\documentclass{mem}
\usepackage{natbib}\usepackage{txfonts}\usepackage{balance}
\usepackage{graphicx}
\usepackage[a4paper]{hyperref}
\idline{0}{001}
\begin{document}
\def\teff{$T\rm_{eff }$}
\def\kms{$\mathrm {km s}^{-1}$}
\def\kmskpc{km s$^{-1}$ kpc$^{-1}$}
\def\deg{$^{\circ}$}

\title{Dynamically possible pattern speeds of double bars}
\subtitle{}
\author{Witold Maciejewski}
\offprints{W. Maciejewski}
\institute{Astrophysics Research Institute, Liverpool John Moores University,
United Kingdom}
\authorrunning{Maciejewski }
\titlerunning{Pattern speeds of double bars}
\abstract{The method to study oscillating potentials of double bars, based on
invariant loops, is introduced here in a new way, intended to be more 
intelligible. Using this method, I show how the orbital structure of a 
double-barred galaxy (nested bars) changes with the variation of nuclear 
bar's pattern speed. Not all pattern speeds are allowed when the inner bar 
rotates in the same direction as the outer bar. Below certain minimum pattern
speed orbital support for the inner bar abruptly disappears, while
high values of this speed lead to loops that are increasingly
round. For values between these two extremes, loops supporting the inner bar
extend further out as its pattern speed decreases, and they become more 
eccentric and pulsate more. These findings do not apply to
counter-rotating inner bars.
}
\maketitle{}

\section{Introduction}
Double-barred galaxies are barred galaxies, where a second, smaller bar is
nested inside the larger bar (see Erwin, this volume, for the review).
Two independent surveys (Erwin \& Sparke 2002; Laine et al. 2002) indicate
that up to 30\% of barred galaxies host nested bars, but cross-correlation
of these samples implies that this percentage may be lower (Moiseev,
this volume). Observed random orientation of the two bars in double-barred
galaxies indicates that the bars may rotate independently. This was confirmed
for NGC 2950, where Tremaine-Weinberg integrals are inconsistent with a single
rotating pattern (Corsini et al. 2003).

Our understanding of the dynamics of barred galaxies strongly relies on studies
of periodic orbits. Maciejewski \& Sparke (1997, 2000) showed that similar
studies can be pursued for double bars. Closed periodic orbits in a single bar
correspond to double-frequency orbits in double bars (Maciejewski \& 
Athanassoula 2007, 2008). These orbits can be studied through their maps
called loops or invariant loops. In Section 2, we show the 
benefit of studying double-frequency orbits with this method. In Section 3,
we apply this method to study how the structure of the inner bar depends
on its pattern speed, and whether all pattern speeds of that bar are 
dynamically possible.

\section{Why study double bars with loops}
In orbital studies one assumes the form of the potential (which can vary
with time), and calculates orbits in this potential. A single bar is usually
treated as a fixed, rigidly rotating potential. 
Stable closed periodic orbits form the backbone of the bar, and
their shapes and extent provide information about the structure of the
bar. However, these orbits close only in the frame rotating with the bar;
in other frames they have a rosette-like appearance (Fig.~1), which no longer
displays the information that the orbit can give about the structure of the
bar. Moreover, if we want to relate the appearance of the orbit to the 
structure of the bar, we should study this orbit in a frame in which the
potential does not change (is stationary). Otherwise every moment on the 
orbit would correspond to a different shape of the potential, and one would
not be able to relate the shape of the orbit to the shape of the potential.

\begin{figure}[!h]
\resizebox{\hsize}{!}{\includegraphics[clip=true]{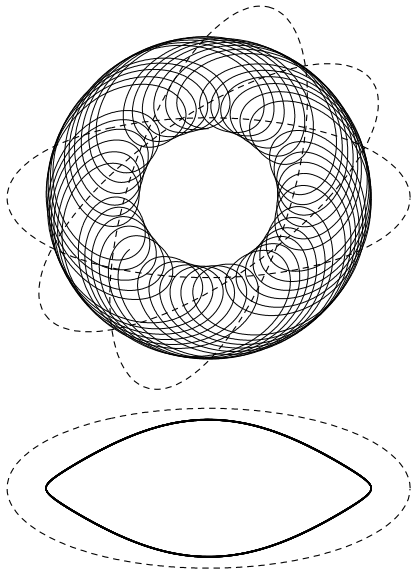}}
\vspace{-20mm}
\caption{\footnotesize Stable closed periodic orbit in a single bar (an $x_1$
orbit) in the frame where the bar rotates (upper panel) and in the frame
rotating with the bar (lower panel). The dashed line outlines the bar.
Only in the lower panel the potential does not change with time (is 
stationary).}
\label{f1}
\end{figure}

Maciejewski \& Athanassoula (2007) showed that 
closed periodic orbits in a single bar correspond to double-frequency
orbits in double bars. This makes orbital studies of double bars more
complicated, because double-frequency orbits do not close. Moreover, if 
the two bars rotate independently, then the potential changes with time, and
relating the shape of the orbit to the shape of the potential, as we could
do in the case of a single bar above, is no longer possible. However, as we 
explain below, we can still relate any particular shape that the periodically
changing potential of double bars takes to a sample of points from the orbit 
taken at moments when the potential has this given shape.

\begin{figure*}[]
\resizebox{\hsize}{!}{\includegraphics[clip=true]{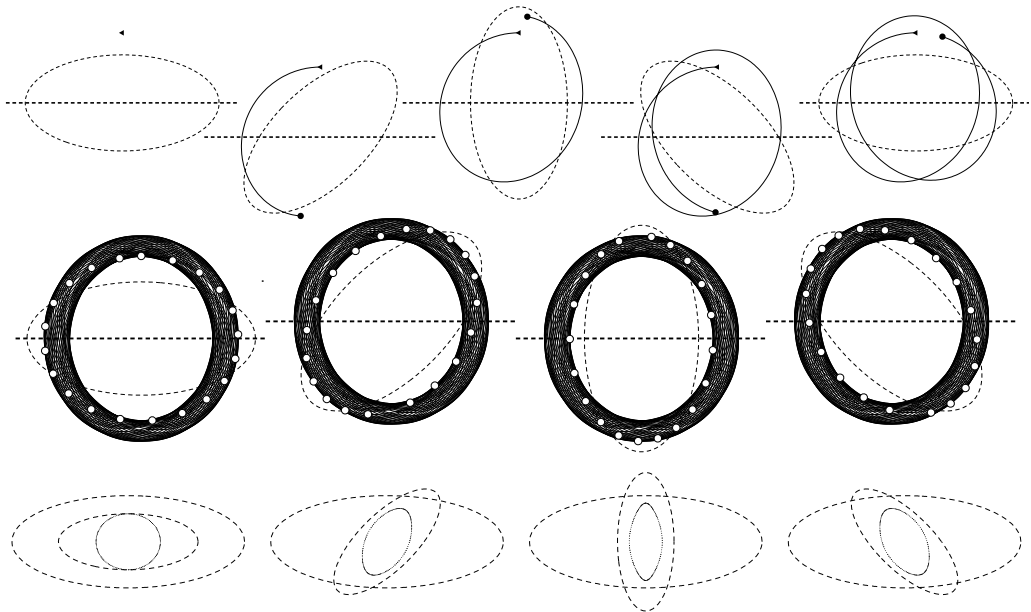}}
\vspace{-115mm}
\caption{
\footnotesize {\bf Top:} Double-frequency orbit in double bars followed from
one to the next alignment of the bars. Triangle marks the starting point and 
the direction of the orbit. Dashed straight horizontal line is drawn along 
the major axis of the outer bar, with which the frame rotates. The inner bar
is outlined with the dashed line. Round dots mark the point on the orbit 
at the relative position of the bars outlined. {\bf Middle:} Same as in the
top row, but for the orbit followed for 20 alignments of the bars. 
{\bf Bottom:} Points on a double-frequency orbit selected only at the moments 
when the relative position of the bars is as outlined with dashed lines. These
points constitute the loop.
}
\label{f2}
\end{figure*}

In the top row of Fig.~2, an example double-frequency orbit in double bars 
is followed as the bars rotate through each other,
and drawn in the frame in which the outer bar remains horizontal (both the 
rotation of the inner bar and the motion on the orbit are counterclockwise).
In a sequence of panels from left to right, the orbit starts
when the bars are aligned at the location marked by a triangle, 
develops as the bars get out of alignment, until they align 
again. In each panel, the relative orientation of the bars is shown for
the moment when the position on the orbit is marked by the round dot. The
crucial observation that underlies the method presented here is that out of
the segment of the orbit presented in each panel, {\it only this dot is
relevant to the shape of the potential shown there}. All other points on 
the presented segment of the orbit are relevant to other shapes of the potential
at other times. As the bars re-align (right-most panel),
the orbit does not close, but out of its segment drawn there, two points are
now relevant to the shape of the potential presented there: the triangle and
the round dot.

Since double-frequency orbits do not close, they contain infinite sets of 
points relevant to any shape that the oscillating potential can take. In the
middle row of Fig.~2, we show a segment of the same double-frequency orbit as
in the top row, but now followed for 20 alignments of the bars. At each
panel, round dots mark points sampled from this orbit at the moments when the
potential takes the shape outlined in a given panel. These points appear to
fall on a closed curve, as confirmed in the bottom row of Fig.~2, where for
clarity we omitted the double-frequency orbit, and only plotted points 
sampled from its segment spanning 60 alignments of the bars. These points
constitute the loop: if a particle is on a non-closing double-frequency orbit,
then at any given shape of the periodically changing potential, it will be
located somewhere on the loop. Thus representing double-frequency orbits with
loops allows us to relate them to any instantaneous shape of the potential,
even though the potential changes with time and the orbit does not close.

It is important to stress here that loop is not an orbit, but it is a 
representation of the orbit, or a sample of relevant points from this orbit.
Loops are unrelated to loop orbits with whom they are sometimes confused.
Orbital analysis is much faster than constructing $N$-body models of double
bars, and since parameters of the bars can be changed arbitrarily, it can
explore any range of parameters of double bars. Maciejewski \& Athanassoula 
(2008) analyzed extent of double-frequency orbits in 23 models of double bars,
whose parameters were varied. They noticed that the trapping of trajectories
around double-frequency orbits strongly depends on the inner bar's pattern
speed. The phase-space volume occupied by trapped orbits monotonically 
increases 
with that pattern speed, accompanied by decreasing chaos. This is contrary to 
previous expectations that resonant coupling between rotating patterns should
minimize chaos (Sygnet et al. 1988). 

\begin{figure*}
\vspace{-14mm}
\resizebox{\hsize}{!}{\includegraphics[clip=true]{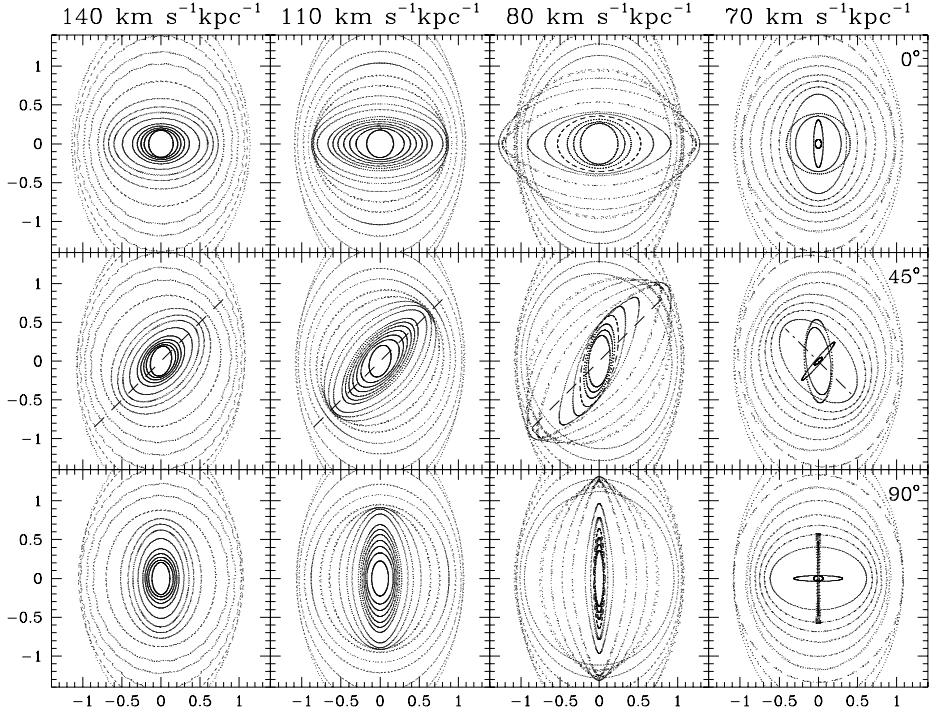}}
\vspace{-29mm}
\caption{
\footnotesize Representative $x_2$ loops (i.e. loops that correspond to the $x_2$
orbits in a single outer bar) for four orbital models of double bars from Maciejewski 
\& Small (in prep.), drawn in a frame in which the outer bar remains 
horizontal. Each column presents a different model, with the pattern speed 
of the inner bar in that model given at the top of the column.
In the top row, the loops are drawn for the moment when the angle 
between the bars is 0\deg, in the middle row when it is 45\deg, while in the
bottom row when this angle is 90\deg\ (as marked in the right-hand column).
In the middle row, the major axis of the inner bar in the imposed 
gravitational potential is drawn with a dashed line, except for the rightmost 
panel, where the minor axis of the inner bar is drawn with a dash-dot line.
Units on axes are in kpc.
}
\label{f3}
\end{figure*}

\section{How orbital support for inner bar changes with its pattern speed}
In a (singly) barred galaxy with an inner Lindblad resonance (ILR), there are 
two major orbital families: the $x_1$ orbits, elongated along the major axis of 
the bar, and the $x_2$ orbits, elongated perpendicularly to it. If within the
ILR, there is another, independently rotating secondary bar (a double-barred 
galaxy), then the loops corresponding to the $x_1$ orbits in the outer bar 
(i.e. the $x_1$ loops) remain elongated with that bar (Maciejewski \& Sparke 
2000). If the secondary bar rotates in the same direction as the outer bar, 
then among the $x_2$
loops (defined as those that correspond to the $x_2$ orbits in the outer bar)
the outer ones remain perpendicular to the outer bar, but the inner ones align
with the inner bar. {\it Note that in this notation, when the two bars rotate 
in the same direction, loops that support the inner bar are the $x_2$ loops -- 
they correspond to the $x_1$ orbits in the inner bar.} When the two bars rotate
in opposite directions, loops that support the inner bar originate from the $x_4$
orbits of the outer bar (Maciejewski 2008).

Maciejewski \& Small (in prep.) studied the $x_2$ loops in seven models of
double bars that rotate in the same direction with the pattern speed of the 
inner bar being set at 70 to 140 \kmskpc, depending on the model. They analyzed 
how the orbital support of the inner bar changes with its pattern speed.
Parameters for five models were taken from models 01-05 of Maciejewski \& 
Athanassoula (2008), hence for everything except for the inner bar's pattern 
speed, they have the values of Model 1 from Maciejewski \& Sparke (2000).

Representative $x_2$ loops for four out of seven models analyzed by Maciejewski 
\& Small (in prep.) are shown in Fig.~3. One can immediately notice several trends, 
further quantified by  Maciejewski \& Small:
\begin{enumerate}
\item The orbital model by Maciejewski \& Sparke (2000) indicates that the 
inner bar should end well within its corotation, which was then confirmed by
$N$-body simulations of double bars (Debattista \& Shen 2007). In Fig.~3 we see
that the orbital support for the inner bar extends further out in radius for
lower pattern speeds. However, lower pattern speed means larger corotation
radius, and the ratio of the extent of orbital support of the inner bar to
its corotation radius remains remarkably constant for the models considered 
here. This implies that the inner bar can extend to $(40\pm2)$\% of its 
corotation.
\item However, one can notice that for orbits that support outer parts of the 
inner bar when its pattern speed is 80 \kmskpc, the sampled points are slightly
scattered around the expected closed curves. This reflects the general trend
for lower pattern speeds that double-frequency orbits supporting outer parts
of the inner bar do not trap large volumes of phase-space, hence provide only
limited support for the bar.
\item The inner bar pulsates as it rotates through the outer bar, as found by
Maciejewski \& Sparke (2000) and confirmed by $N$-body simulations. From
Fig.~3 one can see that
as the pattern speed of the inner bar decreases, loops that support it
become more eccentric and pulsate more as the bars rotate through each 
other.
\item As originally noticed by Maciejewski \& Sparke (2000) and confirmed 
by $N$-body simulations, loops that 
support the inner bar overtake the figure of the bar in the imposed 
potential (its major axis drawn with dashed line in Fig.~3) when the bars
get out of alignment. The magnitude of this effect is virtually constant 
at higher pattern speeds: when the angle between the bars in the imposed
potential is 45\deg, loops supporting the inner bar lead the figure of that
bar by 6\deg$\pm$0.5\deg. Also, the loops rotate coherently at higher pattern
speeds, i.e. their major axes are aligned to within a few degrees.
On the other hand, for pattern speeds 90 \kmskpc and lower,
the angle by which the loops lead the bar increases, and the major axes of
the loops are no longer aligned.
\item The pattern speed of the inner bar cannot be arbitrarily low. As can
be seen in the right-hand column of Fig.~3, when this pattern speed drops
from 80 to 70 \kmskpc, loops that support the inner bar are completely
wiped out. They are replaced by loops, whose orientation changes in relation
to the inner bar (represented by four innermost loops), or remains 
perpendicular to that bar, indicating that they may be related to the $x_2$
orbits in the {\it inner} bar. Interestingly, we found no models in which
loops corresponding to the $x_1$ and $x_2$ orbits in the inner bar coexist. As
can be seen in fig.9 of Maciejewski \& Athanassoula (2008), in linear 
approximation loops corresponding to the $x_2$ orbits in the inner bar should
appear already for pattern speeds of 100 \kmskpc\ and below, but they appear
only at 70 \kmskpc, and their appearance is accompanied by vanishing of
orbits that support the inner bar.
\end{enumerate}

In short, a given secondary inner bar in a double-barred galaxy can rotate
at a rate within a limited range. The lower limit is set by an abrupt
destruction of orbits that support it, while the soft upper limit comes from
the bar becoming increasingly rounder and being no longer a distinct dynamical
feature as its pattern speed increases. These limits apply only to double bars
that rotate in the same direction. As shown by Maciejewski (2008), 
in counter-rotating double bars, inner bars are supported by loops corresponding
to a different orbital family ($x_4$), and their pattern speeds may not be 
limited in a similar way.

\section{Conclusions}
Double bars, like single bars, can be studied with orbital analysis, but
since their appearance oscillates in time, for each instantaneous shape of
the system one should sample points on orbits at moments when the system takes
this given shape. The method of invariant loops, proposed by Maciejewski \&
Sparke (1997, 2000) and developed by Maciejewski \& Athanassoula (2007, 2008)
serves this purpose.

Orbital structure of the inner bar in double-barred galaxies, where both bars 
rotate in the same direction, strongly depends on this bar's pattern speed.
At large pattern speeds loops supporting the inner bar trap large volume of 
phase-space, but they build inner bar that is short and round. At small pattern
speeds the inner bar is longer, but it pulsates and accelerates more, and 
the volume of chaotic zones increases. We find  no evidence of minimizing 
chaos at resonant coupling between the two bars.

This work was partially supported by the Polish Committee for 
Scientific Research as a research project 1 P03D 007 26 in the years 
2004--2007. WM acknowledges Academic Fellowship EP/E500587/1
from Research Councils UK.

\bibliographystyle{aa}

\end{document}